# DSDP: A Blind Docking Strategy Accelerated by GPUs

YuPeng Huang,[1,#] Hong Zhang,[1,#] Siyuan Jiang,[1] Dajiong Yue,[2] Xiaohan Lin,[1] Jun Zhang,[3] Yi Qin Gao[1,*]

[1] College of Chemistry and Molecular Engineering, Biomedical Pioneering Innovation Center, Peking University, Beijing, China.

[2] Huawei Building, No.3 Xinxi Road, Haidian District, Beijing, China 100085

[3] Beijing Changping Lab, Yard 28, Science Park Road, Changping, Beijing, China 102200

*E-mail: gaoyq@pku.edu.cn

[#] Contributed equally to this work

## ABSTRACT

Virtual screening, including molecular docking, plays an essential role in drug discovery. Many traditional and machine-learning based methods are available to fulfil the docking task. The traditional docking methods are normally extensively time-consuming, and their performance in blind docking remains to be improved. Although the runtime of docking based on machine learning is significantly decreased, their accuracy is still limited. In this study, we take the advantage of both traditional and machine-learning based methods, and present a method Deep Site and Docking Pose (DSDP) to improve the performance of blind docking. For the traditional blind docking, the entire protein is covered by a cube, and the initial positions of ligands are randomly generated in the cube. In contract, DSDP can predict the binding site of proteins and provide an accurate searching space and initial positions for the further conformational sampling. The docking task of DSDP makes use of the score function and a similar but modified searching strategy of AutoDock Vina, accelerated by implementation in GPUs. We systematically compare its performance with the state-of-the-art methods, including Autodock Vina, GNINA, QuickVina, SMINA, and DiffDock. DSDP reaches a 29.8% top-1 success rate (RMSD < 2 Å) on an unbiased and challenging test dataset with 1.2 s wall-clock computational time per system. Its performances on DUD-E dataset and the time-split PDBBind dataset used in EquiBind, TankBind, and DiffDock are also effective, presenting a 57.2% and 41.8% top-1 success rate with 0.8 s and 1.0 s per system, respectively.

# 1 INTRODUCTION

Molecular docking is a crucial step to generate potential candidates for lead compounds in drug discovery.[1,2] Docking is composed of several steps, for example, binding pocket identification, drug conformations sampling, scoring, and ranking. Generally, the binding pocket is provided by users in re-docking, cross-docking and virtual-screening tasks, with the pocket being identified by the co-crystal structure of the target protein and associated ligands in the experiments. However, with the development of protein structure prediction methods, for example, AlphaFold,[3] ColabFold,[4] and RosettaFold,[5] a fast increasing number of protein structures are generated without information on ligands. Therefore, it is of high demand to perform reliable ligand docking based on protein structures only and without known binding pockets.

Traditionally, the blind docking is regarded as a task of docking around the entire protein, and many traditional docking programs are available for such tasks, for example, Autodock Vina,[6] and Glide.[7] It is of great value to improve the docking speed and accuracy, given that normally a large space should be sampled in limited searching steps. To deal with such a problem, a number of optimized sampling methods were developed, for instance, QuickVina-W,[8] which was developed based on QuickVina 2.[9,10] QuickVina 2 optimized the local search frequency by searching only potentially important spatial points. These spatial points are identified by checking gradients of the scoring function against a thread history before local optimization. QuickVina-W is a program designed for blind docking, and the potentially significant points are identified by examination of the history of the present and other threads. Besides the improvement on the sampling method, another strategy to increase speed and accuracy is to decrease the searching space through an identification of the potential ligand-protein binding pockets. Methods based on both traditional geometrical or machine learning strategies have been developed to recognize the protein pocket.[11,12] The traditional methods have a relatively long history, and have observed the development of various strategies. For example, in FunFOLD[13] and COFACTOR[14], the binding pocket is located by calculations on the similarity between the target and the templets of known pockets. Methods such as Fpocket,[15] on the other hand, are based on an examination of the shape and spatial geometry of the target protein. In another strategy one performs the binding pocket search using designed probes and identifies the pocket by calculating the interaction energy between the probes and protein.[16] In addition to the traditional methods, the strategies based on machine learning began to show high performance for the binding site prediction over the last few years. Among them, P2Rank[17] is a widely used method based on the random forest algorithm, while COACH[18] is trained by the support vector. In these methods based on deep learning, 3D-CNN are often used, as in DeepSite,[19] DeepSurf [20]and PUResNet.[21]

Besides binding site prediction, many studies focused on combining the site recognizing, pose sampling and scoring in one shot to improve the performance of blind docking. EquiBind[22] is a popular method among them, which applies an SE(3)-equivariant geometric deep learning strategy and successfully decreases the runtime of docking to < 1 s per system. In addition, TANKBind,[23] another deep learning-based method using trigonometry-aware neural networks, replaces the expensive sampling by evaluation of the protein-ligand interaction energy landscapes of different blocks of protein, which further improves the performance in docking tasks. Recently, another state-of-the-art approach, DiffDock,[24] was reported which is based on deep learning and treats the docking as a generative task. DiffDock used diffusion generative model to generate conformations and applied a confidence model to estimate the poses. This method enjoys a significant improvement in the docking accuracy, representing a powerful intermediate approach between traditional sampling and one-shot prediction.

The score function, which is commonly used to estimate the confidence of ligand binding poses, is another important factor affecting the accuracy of blind docking. There are four main categories of scoring functions, namely, physics-based, knowledge-based, empirical, and machine-learning based scoring functions. Many efforts have been paid to improve the performance of score functions, for instance, SMINA,[25] GNINA,[26,27] RF-Score,[28] and IGN.[29] Most of these methods are based on linear regression or machine learning, and present a reasonable performance in estimating the interactions between the proteins and ligands. However, most of the machine-learning based strategies are not introduced directly into the molecular docking procedure in the form of the scoring function, but are used to rescore the poses of ligands generated by the traditional sampling methods. Because a high computational cost is required when the network is used to guide the sampling, implementing a rescoring process after the sampling is a common strategy to improve the accuracy of the latter, as in GNINA.

In the present work, to improve the speed and accuracy of blind docking, we developed a method, Deep Site and Docking Pose (DSDP), to combine the advantages of both machine learning and traditional sampling strategies. It predicts the binding site on the protein and provides the potential location of ligands to decrease the searching space for the following binding pose sampling. A similar strategy was used in EquiBind,[22] DiffDock[24] and Uni-Dock[30]. In these protocols, the binding site identification and ligand conformation sampling are treated separately, and only the predicted site center is used in the sampling step by ignoring the shape of the binding pocket. In the present work, the geometrical information of predicted pockets is used to guide the sampling, including the initial positions and searching space of ligands. In our method, the sampling is not restricted to a fixed cube, but to an adaptive space provided by the binding site predicting step, which efficiently reduces the searching space. In addition to the improvement on the accuracy and speed of binding site prediction, GPU acceleration is introduced to further speed up the sampling process. As a mixed method, DSDP provides an integrated docking workflow, which takes advantage of both traditional and machine-learning based methods and is able to perform blind docking with both high accuracy and a high speed.

## 2 METHODS

DSDP is an integrated docking workflow which makes use of machine learning and traditional sampling strategies, the molecular docking pipeline of which is shown in Figure 1.

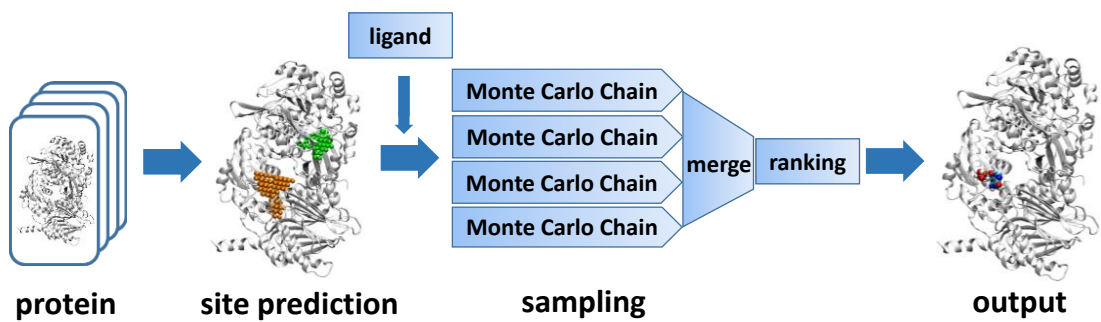

**Figure 1.** Molecular docking pipeline of DSDP.

**Protein-ligand binding site prediction.** The first step of DSDP is the protein-ligand binding site prediction, which is used to provide an accurate searching space for the follow-up binding pose sampling. The workflow of this step is a modified realization of the state-of-the-art binding site prediction method, PUResNet.[21] We introduced several modifications to the original method to further improve the performance of binding site prediction, and more importantly to facilitate the downstream binding pose sampling. In the following part, we provide a brief description of PUResNet as well as a comparison between DSDP and PUResNet. For more details of PUResNet, the readers are referred to the original publication of Chong and coworkers.[21]

PUResNet is a binding site prediction program making use of a 3D-CNN architecture. The protein is represented by $36 \times 36 \times 36$ voxels at its center with 70 Å in each side, and 18 features are introduced to represent each protein atom. The $36 \times 36 \times 36 \times 18$ voxels are used to represent protein structure, which is fed into CNN. The binding site prediction is regarded as a binary segmentation problem, and the binding site can be represented by $36 \times 36 \times 36 \times 1$ voxels at the protein center. The value of the fourth dimension (1 or 0) is used to represent whether this point is a binding site or not. PUResNet combines the classical U-Net and ResNet and it uses the dice loss function. In addition, the training data of PUResNet is a filtered subset of scPDB.[31] It should be noted that the binding site (cavity6.mol2) of scPDB database is generated by the Volsite tool of the IChem program.[32]

The main differences between DSDP and PUResNet are summarized as follows: Firstly, the learning target in DSDP is taken as the position of the explicit ligand atom and not the binding site in scPDB database generated by Volsite as used in PUResNet. Such a choice is made to meet the requirement that an accurate position of ligand is needed in the downstream binding pose sampling. This treatment also allows all databases including the protein and ligand positions to be used as the training data, rather than relying on the scPDB database. Secondly, the 18 chemical features are also modified, with the element Se and S merged to one channel. The newly vacated channel is used to represent the surface of protein, given that the surface of protein is known to play an important role in identifying the protein pockets. Along with this feature, we also provided a surface identifying package (named as surface_tool). To speed up the coordinates reading and feature generation, which in PUResNet are relied on Openbabel[33] and are time-consuming, DSDP reads the protein coordinates by NumPy[34] and loads the features generated by a newly developed toolkit, protein_feature_tool. In protein_feature_tool, we first generated a permanent list by Openbabel outside the inference of site prediction, which includes features of common protein atoms distinguished through residue and atom types. This tool can directly and quickly generate the protein atom feature by searching atom type from the pre-prepared list. Thirdly, we added one extra term to the loss function to the original form of PUResNet, which is the distance between the position of max score and the center of ligand. The weight of this term used in this study is 0.1. We introduced this term to bring the output position of the max score closer to the reference center, given that the location of the sampling center is crucial for the accuracy of searching. The final step in binding site prediction is extracting the pockets from the voxels with different scores. In the original PUResNet, a cutoff of the score value 0.5 and a minimum of pocket size 50 Å$^3$ were used. To exhaustively search for all potential binding pockets, we extracted and clustered the 36 × 36 × 36 × 1 voxels, and selected the top 200 points with high scores. In this way, a series of discrete points describing the binding sites can be provided in the docking stage. After these modifications, we were able to improve the accuracy on binding site prediction (see Figure 6, and further discussion later).

**Binding pose search by traditional sampling methods.** The second component of DSDP is a traditional sampling process accelerated by GPUs, which is similar as AutoDock Vina combined with a number of modifications to the original program. Because the score function of AutoDock Vina was shown to have reached a good balance on accuracy and simplicity during traditional docking, we chose to use the same scoring function in DSDP. Compared to the original Vina docking process, we made three major modifications, as discussed in the following.

Firstly, Vina uses a grid-based method for energy evaluation to reduce the computational cost of protein-ligand interactions.[35] Every vertex of the grid stores a score as a summation over the ligand interactions with protein atoms within the cutoff distance. In this way, the score of a ligand atom can be calculated by trilinear interpolation and its gradient is directly calculated by the partial derivative of the trilinear interpolation. Considering that the gradient is not continuous on the entire interpolation area, we built protein grids that not only contain scores but also their gradients to increase the computational accuracy. The gradient of a ligand atom is also calculated by trilinear interpolation, and therefore is of high accuracy (see the Appendix).

Secondly, a quasi-Newton method, Broyden-Fletcher-Goldfarb-Shanno (BFGS)[36] was used for the local optimization of sampled conformations in Vina. We replaced BFGS by Barzilai-Borwein (BB) method.[37] Unlike BFGS, BB only needs a vector inner product to estimate the step length and does not need to perform line search. This feature makes it well parallelized on GPU. Our testing shows that 100-150 iteration steps are enough to optimize one structure with BB method (see Figure 2).

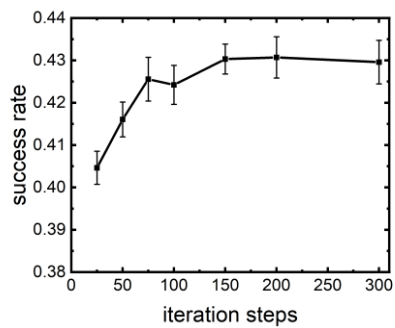

**Figure 2.** Correlation of success rate and the iteration steps.

Thirdly, the searching space of Vina is chosen to be a cube defined by users, and the initial position of the ligand is randomly generated within this box. In the blind docking task, since the entire protein is included in the cube, it is likely that one has to search in a very large sampling space, which can result in a low accuracy. To make full use of information provided by binding site prediction, we used an accurate searching space in the following sampling strategy provided by DSDP. This protocol generates a peripheral $30.0 \times 30.0 \times 30.0$ Å$^3$ cube surrounding the predicted box center for the trilinear interpolation. A union set of balls with 7 Å radius, centered at each discrete points of predicted binding site, is generated (see Figure 3). The intersection of this cube and the union set of balls is taken as the space for binding site search. The initial ligand positions are not fully random but restrained to the searching space. The space of sampling is thus not cubic but has the adaptive shape as described below (see Figure 3). This choice of the searching space makes it resemble closely the shape of cavity and results in a reduced searching volume than the cube.

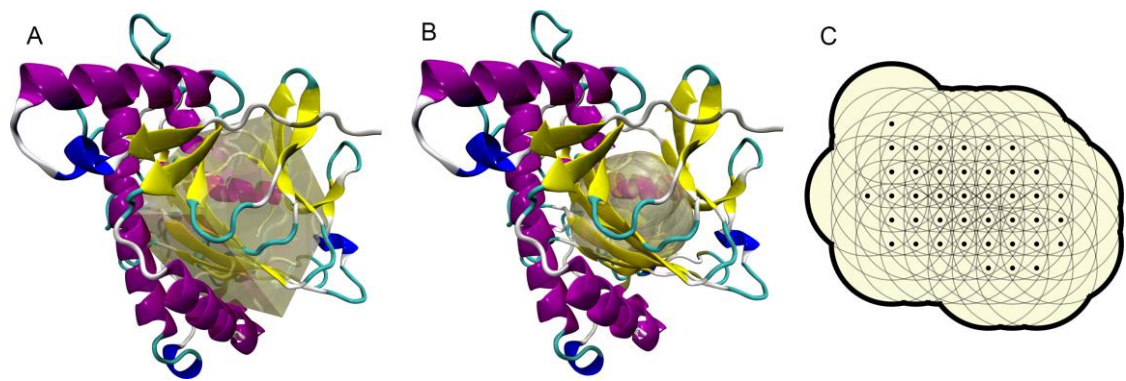

**Figure 3.** The schematic sampling space of (A) traditional cube and (B) adaptive shape. (C) The schematic diagram of the union set of balls with fixed radius with each discrete points of predicted binding site as the center.

As for the parallelization on GPU, to make full use of the parallel computing advantages of GPU, we use a large number of copies (128-2048) and short search steps (20-200) in Monte Carlo searching. This strategy was also used in Vina-GPU[38] and Uni-Dock.[39] The original version of Vina suggests 8 to 64 copies, but each copy would undergo a search of $10^4$-$10^5$ steps. To avoid the performance loss caused by data synchronization, we used the ability of asynchronous concurrent execution provided by CUDA. The initialization and the preparation of copies are run on CPU, which are then packaged and uploaded to GPU. Each copy undergoes Monte Carlo local searching and optimization independently and asynchronously on GPUs. The resulted binding poses are postprocessed in CPU. The workflow of the parallelization on GPUs is shown in Figure 4.

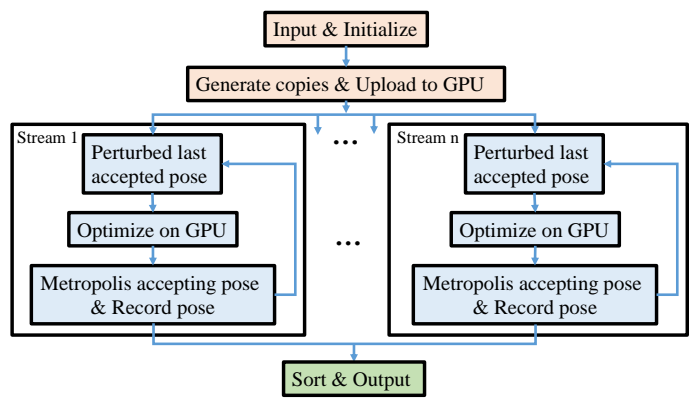

**Figure 4.** Workflow of the parallelization on GPUs. The initialization and the postprocessing of sampled conformations are finished outside the GPUs.

**Data Set preparation.** Considering that many machine-learning based docking methods are trained based on PDBBind database,[40] it would be convenient to compare with other methods using a unified and high quality training dataset. Therefore, we trained the binding site prediction module using the PDBBind v2020 preprocessed in EquiBind.[22] For the test data set, EquiBind, TankBind, and DiffDock all used a new time-split PDBBind dataset. We found that proteins in this latter dataset overlap significantly with the PDBBind v2020 preprocessed dataset (training dataset), which could potentially introduce a bias to the results. To test and compare our method with other state-of-the-art methods in a more rigorous way, we preprocessed a new and clean dataset to estimate the performance of methods. This latest test dataset is a subset of scPDB database without duplication, in which proteins and ligands do not overlap with PDBBind v2020 (see Figure 5). We unified the proteins and ligands by the Uniprot ID and SMILES, respectively. A dataset cleaning process was carried out after we obtained the final complex list. In the preparation of structures, water and metal ions were removed from proteins and ligands. The protein chains were then selected if any atom of them is within a 10 Å radius of any ligand atom. Openbabel,[33] RDkit,[41] Reduce,[42] and ChimeraX[43] were used in the dataset cleaning process. The DUD-E dataset and the time-split PDBBind dataset used in EquiBind, TankBind, and DiffDock were also used to test the performance of DSDP, and the results are presented in the Appendix. The speed of DSDP is the fastest for all three datasets tested, and its success rate is higher than the other methods except for GNINA using 64 sampling copies and DiffDock in DUD-E dataset.

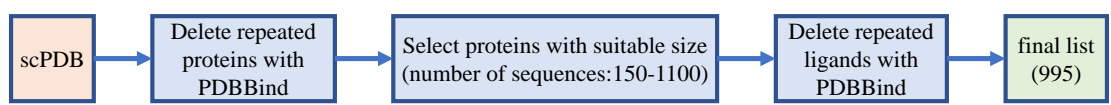

**Figure 5.** Workflow of filtering the final test dataset from original scPDB database. As a result of this procedure, we constructed a dataset containing 995 ligand-protein complexes.

**Baseline Setup.** In order to estimate the accuracy of binding site prediction of DSDP, we compared its binding site prediction function with that of P2Rank and PUResNet. P2Rank and PUResNet programs were run using the provided models provided by the authors. To estimate the sampling performance of DSDP, we compared the docking speed and accuracy with AutoDock Vina, QuickVina-W (QuickVina2), GNINA, SMINA and DiffDock. We followed the protocol suggested by QuickVina and QuickVina-W and QuickVina2 are used to do blind docking and redocking, respectively. For the traditional docking programs AutoDock Vina, QuickVina-W (QuickVina2), GNINA, and SMINA, we found that the parameter exhaustiveness affects strongly the runtime and performance. Therefore, we used two different values, exhaustiveness = 8 and exhaustiveness = 64, to estimate its effect. Three parallel redocking and blind docking calculations were performed in the present work. DiffDock was performed using the default hyperparameters but with a batch size of 1. The heavy-atom root mean squared deviation (RMSD) was calculated to estimate the accuracy of docking. To unify the hardware and exhaust the best parallelism performance of Vina-based methods, we used 64 CPUs (Intel Xeon Platinum 8358) for AutoDock Vina, QuickVina, and SMINA in the redocking and blind docking tasks. It should be mentioned that 64 CPUs are not fully utilized when the exhaustiveness is set to 8. For the GPU-based methods, namely, GNINA, DSDP and DiffDock, we used NVIDIA GeForce RTX 2080 SUPER GPU (AMD Ryzen 7 2700X Eight-Core Processor).

## 3 RESULTS AND DISCUSSION

**Protein-ligand binding site prediction.** To evaluate the performance of protein-ligand binding site prediction of DSDP, we compared our method with PUResNet and P2Rank. We mainly used two parameters to evaluate the performance of our model, which are highly related to the downstream sampling, namely, the distance between predicted binding site center to the center of actual ligand (DCC), and the volume coverage rate (VCR). The former quantity determines the sampling position of ligands, and the latter is explained as follows. In the molecular docking process, a restrained wall is needed to ensure that the docking is only performed in a restricted spatial region. At the same time, the searching space needs to cover the entire ligand, namely, the volume coverage rate of the searching space to the actual ligand, should be 1. As shown in Figure 6A, we find that DSDP has a higher success rate of DCC than the other methods. This result indicates that DSDP provides a more accurate location of the sampling center. We then calculated the VCR using cubes of different side lengths, and the cube center is the predicted binding site center. As depicted in Figure 6B, DSDP again shows the highest success rate of VCR among all methods tested. Considering that the searching space of DSDP is an adaptive shape rather than a cube, we compared the size of volume based on the adaptive volume and the $22.5 \times 22.5 \times 22.5$ Å$^3$ cube, which is widely used in re-docking task.[6, 39] We found that the adaptive shape is considerably smaller than the cube (red line in Figure 6C). This comparison indicates that using the adaptive shape in the downstream sampling can effectively reduce the search over futile space.

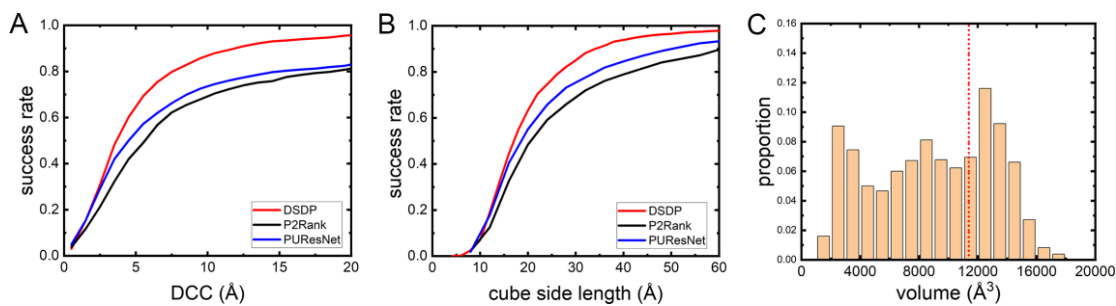

**Figure 6.** Estimation of (A) DCC and (B) VCR of DSDP, PUResNet, and P2Rank. P2Rank provides many pockets with different scores and we only use the top-1 output. PUResNet and DSDP use all outputs (the average output number of pockets no more than 2 per system). The definition of success in DCC is that the distance between any pocket and the ligand center is within a specific value. The definition of success in VCR is that the cube with specific length can cover the whole ligand. (C) Volume proportion of the adaptive shape formed by the union set of balls with 7 Å radius, and the red dotted line is the volume of $22.5 \times 22.5 \times 22.5$ Å$^3$ cube.

**Re-docking in the known site.** DSDP is not only used for blind docking, but also the re-docking task. Before the blind docking test, in the following we verify first the docking accuracy based on tests on cases with known binding sites. The known binding site is defined by the position of the co-crystalized ligand, and the box size is obtained by adding 4 Å along the negative and positive directions for minimum and maximum of the x, y, and z coordinates of the ligand, respectively. As listed in Table 1, we performed ligand docking using AutoDock Vina, QuickVina2, GNINA, SMINA with different number of copies (the parameter exhaustiveness), as well as DSDP. It should be mentioned that the number of sampling steps of Monte Carlo in the Vina-based methods varies with the number of atoms and degree of freedom of the ligands, which is around $10^4$ to $10^5$ steps. QuickVina2 optimized the searching strategy to reduce the number of required search steps. In practice, the numbers of copies and searching steps are balanced to optimize jointly the computational speed and accuracy. For Vina, the accuracy changes slightly but the runtime increases significantly when one increases the sampling copies from 8 to 64. Such an observation indicates that the increase of the sampling steps does not bring much benefits to the redocking task, because the sampling space is limited. The runtime of QuickVina2 is shorter than the other Vina-based methods. On the other hand, GNINA enjoys a higher accuracy than the methods mentioned above when the same sampling parameters are used, resulted from a rescoring process after the sampling. Rescoring helps extract the lower RMSD conformation of ligand from the top 50 poses generated by the traditional sampling method SMINA independent of the sampling process. For the implementation of DSDP, we found that satisfactory performance is obtained when the numbers of copies and searching steps are set to 384 and 40, respectively, which are far fewer than the sampling numbers required by Vina-based methods mentioned above. Although the sampling number is small, the accuracy of DSDP in re-docking task (the RMSD of top-1 pose within 2 Å) matches the Vina-based methods using 8 copies. Notably, the runtime of DSDP (384 ×

40) is much less than the other methods, showing a speed-up of ~ 100 times compared to the original Vina using 64 CPUs with the exhaustiveness=64.

**Table 1 Redocking of test dataset.** The runtime of DSDP does not include the time of protein binding pocket prediction in the redocking task. *64 CPUs were used for Vina, QuickVina2, and SMINA. #GPU was used in GNINA and DSDP.

| Methods | RMSD < 2 Å (%) | RMSD < 5 Å (%) | Time (s) | Sampling |
|---|---|---|---|---|
| Vina | 43.2 ± 0.4 | 70.6 ± 0.8 | 36.0* | exhaustiveness=8 |
| | 44.3 ± 0.1 | 70.9 ± 0.4 | 70.6* | exhaustiveness=64 |
| QuickVina2 | 41.3 ± 0.5 | 68.8 ± 0.3 | 6.9* | exhaustiveness=8 |
| | 43.2 ± 0.5 | 69.8 ± 0.5 | 7.7* | exhaustiveness=64 |
| SMINA | 42.2 ± 0.1 | 69.8 ± 0.4 | 18.7* | exhaustiveness=8 |
| | 44.1 ± 0.2 | 70.7 ± 0.4 | 22.9* | exhaustiveness=64 |
| GNINA | 54.0 ± 0.1 | 84.1 ± 1.0 | 23.0# | exhaustiveness=8 |
| | 55.1 ± 0.4 | 85.0 ± 0.4 | 100.0# | exhaustiveness=64 |
| DSDP | 39.1 ± 0.2 | 74.3 ± 0.6 | 0.33# | 128 × 20 |
| | 42.5 ± 0.3 | 74.6 ± 0.4 | 0.55# | 384 × 40 |
| | 42.5 ± 0.6 | 73.6 ± 0.6 | 0.64# | 512 × 40 |
| | 43.5 ± 0.7 | 72.4 ± 0.2 | 1.71# | 512 × 200 |
| | 43.7 ± 0.6 | 71.2 ± 1.0 | 3.28# | 1024 × 200 |
| | 43.8 ± 0.3 | 71.1 ± 0.3 | 6.32# | 2048 × 200 |

**Blind docking.** To evaluate the performance of DSDP on blind docking, we also used our preprocessed dataset as the test set. As shown in Table 2, we found that the number of sampling copies affects the accuracy of blind docking in traditional methods. For example, AutoDock Vina presents a 12.9% success rate (RMSD < 2 Å for the top-1 poses) with 8 sampling copies, while this rate increases to 24.6% when the parameter exhaustiveness is increased to 64. This behavior of the model can also be found in other traditional methods. However, further increasing this parameter is not recommended due to the fast increase of the demands on hardware and runtime. AutoDock Vina and SMINA yield similar results when the parameters of configuration used for them are the same as expected, since they share the same sampling strategy. Compared to these methods, QuickVina-W uses fewer searching steps but has a slightly decreased success rate, namely 10.6% and 21.2% for 8 and 64 sampling copies, respectively. GNINA again shows the best accuracy, benefitting from the rescoring process. The performance of DiffDock on this new dataset is less optimal compared with that is seen in the original paper. To compare the influence of the choice of test sets, we used the same parameters of DiffDock in the present work to study the test set (363 complexes) used in the original paper.[24] We obtained a 36.5% success rate (see Table A1 in Appendix), which reproduces the results of Ref.[24] The success rate of DSDP on the 363 dataset is 41.8%, which is higher than that of DiffDock. The comparison on the DUD-E dataset between different methods is also provided in Table A2, and the success rate on this dataset is higher than the other datasets. These results indicate that our test set (995 complexes) represents a more challenging task than the 363 complexes and DUD-E dataset. In addition, we examined the effects of the searching space shape (cube and adaptive shape) on the speed and accuracy of blind docking using DSDP. The size of the search cube is 22.5 × 22.5 × 22.5 Å$^3$ using the predicted binding site as the center, and the adaptive shape is obtained by a union set of balls with 7 Å radius (see more details in Methods). As shown in Table 2, we found that the docking accuracy of adaptive shapes is slightly higher than that of cubes (22.5 × 22.5 × 22.5 Å$^3$). In addition, the runtime based on the adaptive shape (1.20 s) is shorter than that for the cubic one (1.26 s). Such a difference can be readily rationalized since the former samples a more relevant space than the latter. We found that although the combination of searching steps and copies for DSDP is far less than Vina-based method including even QuickVina-W, the success rate of the former (29.8%) is significantly higher than the others except for GNINA using 64 sampling copies. This result indicates that the binding site prediction used in DSDP can not only pinpoint the sampling position but also decrease the sampling space.

**Table 2 Blind docking of test dataset.** The runtime of DSDP includes the time of protein binding pocket prediction in the blind docking task. [*]64 CPUs were used for Vina, QuickVina2, and SMINA. [#]GPU was used in GNINA, DiffDock and DSDP.

| Methods | RMSD < 2 Å (%) | RMSD < 5 Å (%) | Time (s) | Sampling |
| --- | --- | --- | --- | --- |
| Vina | 12.9 ± 0.5 | 26.8 ± 1.7 | 56.5[*] | exhaustiveness=8 |
| | 24.6 ± 0.3 | 39.4 ± 0.4 | 93.9[*] | exhaustiveness=64 |
| QuickVina-W | 10.6 ± 0.9 | 23.0 ± 1.2 | 11.5[*] | exhaustiveness=8 |
| | 21.2 ± 0.1 | 35.8 ± 0.2 | 13.5[*] | exhaustiveness=64 |
| SMINA | 12.4 ± 0.3 | 26.1 ± 1.8 | 22.3[*] | exhaustiveness=8 |
| | 23.9 ± 0.5 | 39.1 ± 0.4 | 24.0[*] | exhaustiveness=64 |
| GNINA | 25.6 ± 1.5 | 39.2 ± 2.0 | 33.5[#] | exhaustiveness=8 |
| | 37.7 ± 0.7 | 53.7 ± 0.1 | 135.1[#] | exhaustiveness=64 |
| DiffDock | 17.6 ± 0.6 | 43.5 ± 0.2 | 86[#] | - |
| DSDP (cube) | 28.4 ± 0.6 | 46.2 ± 0.3 | 1.26[#] | 384 × 40 |
| DSDP (shape) | 29.8 ± 0.3 | 47.9 ± 0.1 | 1.20[#] | 384 × 40 |

## 4 CONCLUSION

To improve the performance of molecular docking, many machine-learning based methods have been introduced into the docking tasks, leading to significant successes in the prediction of protein binding site and estimation of protein-ligand interaction. However, the ligand pose sampling is still a challenge for these methods, partly because the available training data of ligand binding poses is highly limited. On the other hand, the traditional sampling methods have observed great development over the last twenty years, although the calculation efficiency is normally low. In the present work, we try to take advantages of both machine-learning based binding site prediction and traditional sampling methods, and compile an accelerated two task-in-one program on GPU. The method is named DSDP. DSDP is an integrated docking program developed for blind docking, which can also be used for redocking and virtual-screening. To compare DSDP with other state-of-the-art methods, we preprocessed a new and rigorous dataset that has no ligand or protein overlap with the training dataset (PDBBind v2020). DSDP was shown to reach a 29.8% top-1 success rate (RMSD < 2 Å) on this dataset with a computational time about 1.20 s per system, thus enjoying both high calculation efficiency and high accuracy. DSDP makes full usage of the output of binding site prediction, in which the sampling space and the initial positions of ligands are guided to adapt to the pocket. In addition, the binding pose sampling is significantly accelerated by implementation on GPUs, resulting in a speed-up of about 100 times compared to the original Vina using 64 CPUs with the exhaustiveness=64 for both redocking and blind docking. Therefore, this integrated program has the benefits of both traditional and machine-learning based methods, and we hope that it will prove to be a powerful program for the future molecular docking studies.

**ACKNOWLEDGMENTS**

The authors thank National Natural Science Foundation of China (22050003 and 92053202).

**REFERENCE**

(1)  Vamathevan, J.; Clark, D.; Czodrowski, P.; Dunham, I.; Ferran, E.; Lee, G.; Li, B.; Madabhushi, A.; Shah, P.; Spitzer, M.; Zhao, S. Applications of Machine Learning in Drug Discovery and Development. *Nat. Rev. Drug Discov.* **2019**, *18* (6), 463–477.

(2)  Saikia, S.; Bordoloi, M. Molecular Docking: Challenges, Advances and Its Use in Drug Discovery Perspective. *Curr. Drug Targets* **2018**, *20* (5), 501–521.

(3)  Jumper, J.; Evans, R.; Pritzel, A.; Green, T.; Figurnov, M.; Ronneberger, O.; Tunyasuvunakool, K.; Bates, R.; Žídek, A.; Potapenko, A.; Bridgland, A.; Meyer, C.; Kohl, S. A. A.; Ballard, A. J.; Cowie, A.; Romera-Paredes, B.; Nikolov, S.; Jain, R.; Adler, J.; Back, T.; Petersen, S.; Reiman, D.; Clancy, E.; Zielinski, M.; Steinegger, M.; Pacholska, M.; Berghammer, T.; Bodenstein, S.; Silver, D.; Vinyals, O.; Senior, A. W.; Kavukcuoglu, K.; Kohli, P.; Hassabis, D. Highly Accurate Protein Structure Prediction with AlphaFold. *Nature* **2021**, *596* (7873), 583–589.

(4)  Mirdita, M.; Schütze, K.; Moriwaki, Y.; Heo, L.; Ovchinnikov, S.; Steinegger, M. ColabFold: Making Protein Folding Accessible to All. *Nat. Methods* **2022**, *19* (6), 679–682.

(5)  Baek, M.; DiMaio, F.; Anishchenko, I.; Dauparas, J.; Ovchinnikov, S.; Lee, G. R.; Wang, J.; Cong, Q.; Kinch, L. N.; Dustin Schaeffer, R.; Millán, C.; Park, H.; Adams, C.; Glassman, C. R.; DeGiovanni, A.; Pereira, J. H.; Rodrigues, A. V.; Van Dijk, A. A.; Ebrecht, A. C.; Opperman, D. J.; Sagmeister, T.; Buhlheller, C.; Pavkov-Keller, T.; Rathinaswamy, M. K.; Dalwadi, U.; Yip, C. K.; Burke, J. E.; Christopher Garcia, K.; Grishin, N. V.; Adams, P. D.; Read, R. J.; Baker, D. Accurate Prediction of Protein Structures and Interactions Using a Three-Track Neural Network. *Science (80-. ).* **2021**, *373* (6557), 871–876.

(6)  Trott, O.; Olson, A. J. AutoDock Vina: Improving the Speed and Accuracy of Docking with a New Scoring Function, Efficient Optimization, and Multithreading. *J. Comput. Chem.* **2010**, *31* (2), 455–461.

(7) Friesner, R. A.; Banks, J. L.; Murphy, R. B.; Halgren, T. A.; Klicic, J. J.; Mainz, D. T.; Repasky, M. P.; Knoll, E. H.; Shelley, M.; Perry, J. K.; Shaw, D. E.; Francis, P.; Shenkin, P. S. Glide: A New Approach for Rapid, Accurate Docking and Scoring. 1. Method and Assessment of Docking Accuracy. *J. Med. Chem.* **2004**, *47* (7), 1739–1749.

(8) Hassan, N. M.; Alhossary, A. A.; Mu, Y.; Kwoh, C. K. Protein-Ligand Blind Docking Using QuickVina-W with Inter-Process Spatio-Temporal Integration. *Sci. Rep.* **2017**, *7* (1), 1–13.

(9) Alhossary, A.; Handoko, S. D.; Mu, Y.; Kwoh, C. K. Fast, Accurate, and Reliable Molecular Docking with QuickVina 2. *Bioinformatics* **2015**, *31* (13), 2214–2216.

(10) Handoko, S. D.; Ouyang, X.; Su, C. T. T.; Kwoh, C. K.; Ong, Y. S. QuickVina: Accelerating AutoDock Vina Using Gradient-Based Heuristics for Global Optimization. *IEEE/ACM Trans. Comput. Biol. Bioinforma.* **2012**, *9* (5), 1266–1272.

(11) Macari, G.; Toti, D.; Polticelli, F. Computational Methods and Tools for Binding Site Recognition between Proteins and Small Molecules: From Classical Geometrical Approaches to Modern Machine Learning Strategies. *J. Comput. Aided. Mol. Des.* **2019**, *33* (10), 887–903.

(12) Zhao, J.; Cao, Y.; Zhang, L. Exploring the Computational Methods for Protein-Ligand Binding Site Prediction. *Comput. Struct. Biotechnol. J.* **2020**, *18*, 417–426.

(13) Roche, D. B.; Buenavista, M. T.; McGuffin, L. J. The FunFOLD2 Server for the Prediction of Protein–Ligand Interactions. *Nucleic Acids Res.* **2013**, *41* (W1), W303–W307.

(14) Roy, A.; Yang, J.; Zhang, Y. COFACTOR: An Accurate Comparative Algorithm for Structure-Based Protein Function Annotation. *Nucleic Acids Res.* **2012**, *40* (W1), W471–W477.

(15) Le Guilloux, V.; Schmidtke, P.; Tuffery, P. Fpocket: An Open Source Platform for Ligand Pocket Detection. *BMC Bioinformatics* **2009**, *10* (1), 168.

(16) Tsujikawa, H.; Sato, K.; Wei, C.; Saad, G.; Sumikoshi, K.; Nakamura, S.; Terada, T.; Shimizu, K. Development of a Protein–Ligand-Binding Site Prediction Method Based on Interaction Energy and Sequence Conservation. *J. Struct. Funct. Genomics* **2016**, *17* (2), 39–49.

(17) Krivák, R.; Hoksza, D. P2Rank: Machine Learning Based Tool for Rapid and Accurate Prediction of Ligand Binding Sites from Protein Structure. *J. Cheminform.* **2018**, *10* (1), 39.

(18) Yang, J.; Roy, A.; Zhang, Y. Protein–Ligand Binding Site Recognition Using Complementary Binding-Specific Substructure Comparison and Sequence Profile Alignment. *Bioinformatics* **2013**, *29* (20), 2588–2595.

(19) Jiménez, J.; Doerr, S.; Martínez-Rosell, G.; Rose, A. S.; De Fabritiis, G. DeepSite: Protein-Binding Site Predictor Using 3D-Convolutional Neural Networks. *Bioinformatics* **2017**, *33* (19), 3036–3042.

(20) Mylonas, S. K.; Axenopoulos, A.; Daras, P. DeepSurf: A Surface-Based Deep Learning Approach for the Prediction of Ligand Binding Sites on Proteins. *Bioinformatics* **2021**, *37* (12), 1681–1690.

(21) Kandel, J.; Tayara, H.; Chong, K. T. PUResNet: Prediction of Protein-Ligand Binding Sites Using Deep Residual Neural Network. *J. Cheminform.* **2021**, *13* (1), 1–14.

(22) Stärk, H.; Ganea, O.-E.; Pattanaik, L.; Barzilay, R.; Jaakkola, T. EquiBind: Geometric Deep Learning for Drug Binding Structure Prediction. **2022**.

(23) Lu, W. TANKBind : Trigonometry-Aware Neural NetworKs for Drug-Protein Binding Structure Prediction ArXiv : Submit / 4332744 [ Cs . LG ] 31 May 2022. *bioRxiv* **2022**.

(24) Corso, G.; Stärk, H.; Jing, B.; Barzilay, R.; Jaakkola, T. DiffDock: Diffusion Steps, Twists, and Turns for Molecular Docking. **2022**, arXiv:2210.01776v1.

(25) Koes, D. R.; Baumgartner, M. P.; Camacho, C. J. Lessons Learned in Empirical Scoring with Smina from the CSAR 2011 Benchmarking Exercise. *J. Chem. Inf. Model.* **2013**, *53* (8), 1893–1904.

(26) McNutt, A. T.; Francoeur, P.; Aggarwal, R.; Masuda, T.; Meli, R.; Ragoza, M.; Sunseri, J.; Koes, D. R. GNINA 1.0: Molecular Docking with Deep Learning. *J. Cheminform.* **2021**, *13* (1), 1–20.

(27) Francoeur, P. G.; Masuda, T.; Sunseri, J.; Jia, A.; Iovanisci, R. B.; Snyder, I.; Koes, D. R. Three-Dimensional Convolutional Neural Networks and a Crossdocked Data Set for Structure-Based Drug Design. *J. Chem. Inf. Model.* **2020**, *60* (9), 4200–4215.

(28) Ballester, P. J.; Mitchell, J. B. O. A Machine Learning Approach to Predicting Protein-Ligand Binding Affinity with Applications to Molecular Docking. *Bioinformatics* **2010**, *26* (9), 1169–1175.

(29) Jiang, D.; Hsieh, C. Y.; Wu, Z.; Kang, Y.; Wang, J.; Wang, E.; Liao, B.; Shen, C.; Xu, L.; Wu, J.; Cao, D.; Hou, T. InteractionGraphNet: A Novel and Efficient Deep Graph Representation Learning Framework for Accurate Protein-Ligand Interaction Predictions. *J. Med. Chem.* **2021**, *64* (24), 18209–18232.

(30) Yu, Y.; Lu, S.; Gao, Z.; Zheng, H.; Ke, G. Do Deep Learning Models Really Outperform Traditional Approaches in Molecular Docking? **2023**, arXiv:2302.07134.

(31) Desaphy, J.; Bret, G.; Rognan, D.; Kellenberger, E. Sc-PDB: A 3D-Database of Ligandable Binding Sites-10 Years On. *Nucleic Acids Res.* **2015**, *43* (D1), D399–D404.

(32) Desaphy, J.; Azdimousa, K.; Kellenberger, E.; Rognan, D. Comparison and Druggability Prediction of Protein-Ligand Binding Sites from Pharmacophore-Annotated Cavity Shapes. *J. Chem. Inf. Model.* **2012**, *52* (8), 2287–2299.

(33) O'Boyle, N. M.; Banck, M.; James, C. A.; Morley, C.; Vandermeersch, T.; Hutchison, G. R. Open Babel: An Open Chemical Toolbox. *J. Cheminform.* **2011**, *3* (1), 33.

(34) Harris, C. R.; Millman, K. J.; van der Walt, S. J.; Gommers, R.; Virtanen, P.; Cournapeau, D.; Wieser, E.; Taylor, J.; Berg, S.; Smith, N. J.; Kern, R.; Picus, M.; Hoyer, S.; van Kerkwijk, M. H.; Brett, M.; Haldane, A.; del Río, J. F.; Wiebe, M.; Peterson, P.;

Gérard-Marchant, P.; Sheppard, K.; Reddy, T.; Weckesser, W.; Abbasi, H.; Gohlke, C.; Oliphant, T. E. Array Programming with NumPy. *Nature* **2020**, *585* (7825), 357–362.

(35) Morris, G. M.; Goodsell, D. S.; Halliday, R. S.; Huey, R.; Hart, W. E.; Belew, R. K.; Olson, A. J. Automated Docking Using a Lamarckian Genetic Algorithm and an Empirical Binding Free Energy Function. *J. Comput. Chem.* **1998**, *19* (14), 1639–1662.

(36) Nocedal, J.; Wright, S. J. Numerical Optimization; Springer Verlag: Berlin, 1999, Springer Series in Operations Research.

(37) Fletcher, R. On the Barzilai-Borwein Method; 2005.

(38) Tang, S.; Chen, R.; Lin, M.; Lin, Q.; Zhu, Y.; Ding, J.; Hu, H.; Ling, M.; Wu, J. Accelerating AutoDock Vina with GPUs. *Molecules* **2022**, *27* (9), 1–18.

(39) Yu Y, Cai C, Zhu Z, Zheng H. Uni-Dock: A GPU-Accelerated Docking Program Enables Ultra-Large Virtual Screening. ChemRxiv, **2022**, 10.26434/chemrxiv-2022-5t5ts.

(40) Wang, R.; Fang, X.; Lu, Y.; Wang, S. The PDBbind Database: Collection of Binding Affinities for Protein-Ligand Complexes with Known Three-Dimensional Structures. *J. Med. Chem.* **2004**, *47* (12), 2977–2980.

(41) Landrum, G. Rdkit: Open-source cheminformatics software, **2016**.

(42) https://github.com/rlabduke/reduce.

(43) Pettersen, E. F.; Goddard, T. D.; Huang, C. C.; Meng, E. C.; Couch, G. S.; Croll, T. I.; Morris, J. H.; Ferrin, T. E. UCSF ChimeraX: Structure Visualization for Researchers, Educators, and Developers. *Protein Sci.* **2021**, *30* (1), 70–82.

**APPENDIX**

The modification of the interpolation strategy is shown below:

$$s_i(\boldsymbol{r}_i) = \sum_j f_{ij}(\boldsymbol{r}_i, \boldsymbol{r}_j) \tag{1}$$

$$\approx \sum_{l,m,n=0,1} \left( p_x^l (1-p_x)^{1-l} p_y^m (1-p_y)^{1-m} p_z^n (1-p_z)^{1-n} \sum_j f_{ij}(\boldsymbol{m}_{lmn}, \boldsymbol{r}_j) \right) \tag{2}$$

$$p_x = (\boldsymbol{r}_i - \boldsymbol{m}_{000}) \cdot \left(\frac{1}{l_x}, 0, 0\right)$$

$$p_y = (\boldsymbol{r}_i - \boldsymbol{m}_{000}) \cdot \left(0, \frac{1}{l_y}, 0\right)$$

$$p_z = (\boldsymbol{r}_i - \boldsymbol{m}_{000}) \cdot \left(0, 0, \frac{1}{l_z}\right)$$

where $s_i(\boldsymbol{r}_i)$ is the summation of pair interactions for atom *i* at *r* position. Trilinear interpolation was used to approximatively evaluate this score to increase the calculation efficiency (see equation 2). Specifically, let $r_i$ as one point among $m_{000}, ..., m_{111}$, and $s_i(r_i)$ can be obtained by the linear combination of $\sum_j f_{ij}(m_{lmn}, r_j)$ at the eight grid points, where $l_x, l_y, l_z$ are the side lengths of the grid. It can be found that the gradient of this score is not continuous on the entire interpolation area but a piecewise continuous one (see equation 3). To obtain a continuous gradient in the entire interpolation area, the gradients were also calculated by trilinear interpolation (see equation 4).

$$\nabla \left( \sum_{l,m,n=0,1} \left( p_x^l (1-p_x)^{1-l} p_y^m (1-p_y)^{1-m} p_z^n (1-p_z)^{1-n} \sum_j f_{ij}(m_{lmn}, r_j) \right) \right) \quad (3)$$

$$= \begin{pmatrix} \sum_{l,m,n=0,1} \left( \frac{2l-1}{l_x} p_y^m (1-p_y)^{1-m} p_z^n (1-p_z)^{1-n} \sum_j f_{ij}(m_{lmn}, r_j) \right) \\ \sum_{l,m,n=0,1} \left( p_x^l (1-p_x)^{1-l} \frac{2m-1}{l_y} p_z^n (1-p_z)^{1-n} \sum_j f_{ij}(m_{lmn}, r_j) \right) \\ \sum_{l,m,n=0,1} \left( p_x^l (1-p_x)^{1-l} p_y^m (1-p_y)^{1-m} \frac{2n-1}{l_z} \sum_j f_{ij}(m_{lmn}, r_j) \right) \end{pmatrix}$$

$$\nabla_{r_i} s_i(r_i) \approx \sum_{l,m,n=0,1} \left( p_x^l (1-p_x)^{1-l} p_y^m (1-p_y)^{1-m} p_z^n (1-p_z)^{1-n} \sum_j \nabla_{m_{lmn}} f_{ij}(m_{lmn}, r_j) \right) \quad (4)$$

**Table A1 Blind docking of time-split PDBBind dataset.** The runtime of DSDP includes the time of protein binding pocket prediction in the blind docking task. [*]64 CPUs were used for Vina, QuickVina2, and SMINA. [#]GPU was used in GNINA, DiffDock and DSDP.

| Methods | RMSD < 2 Å (%) | RMSD < 5 Å (%) | Time (s) | Sampling |
|---|---|---|---|---|
| **Vina** | 20.5 | 36.3 | 98.8[*] | exhaustiveness=8 |
| | 32.4 | 49.0 | 199.7[*] | exhaustiveness=64 |
| **QuickVina-W** | 15.5 | 28.3 | 8.3[*] | exhaustiveness=8 |
| | 30.6 | 44.9 | 15.2[*] | exhaustiveness=64 |
| **SMINA** | 21.6 | 36.6 | 51.7[*] | exhaustiveness=8 |
| | 36.3 | 52.1 | 99.7[*] | exhaustiveness=64 |
| **GNINA** | 35.5 | 50.4 | 93.1[#] | exhaustiveness=8 |
| | 51.0 | 64.7 | 509.4[#] | exhaustiveness=64 |
| **DiffDock** | 36.5 | 63.2 | 70.4[#] | - |
| **DSDP (shape)** | 41.8 ± 1.8 | 61.3 ± 2.1 | 1.04[#] | 384 × 40 |

**Table A2 Blind docking of DUD-E dataset.** The runtime of DSDP includes the time of protein binding pocket prediction in the blind docking task. *64 CPUs were used for Vina, QuickVina2, and SMINA. #GPU was used in GNINA, DiffDock and DSDP.

| Methods | RMSD < 2 Å (%) | RMSD < 5 Å (%) | Time (s) | Sampling |
|---|---|---|---|---|
| Vina | 19.6 | 30.4 | 21.2* | exhaustiveness=8 |
|  | 43.1 | 59.8 | 53.5* | exhaustiveness=64 |
| QuickVina-W | 15.7 | 24.5 | 6.5* | exhaustiveness=8 |
|  | 42.0 | 55.0 | 11.8* | exhaustiveness=64 |
| SMINA | 16.8 | 28.7 | 11.2* | exhaustiveness=8 |
|  | 41.2 | 52.9 | 21.2* | exhaustiveness=64 |
| GNINA | 52.9 | 67.6 | 20.6# | exhaustiveness=8 |
|  | 71.6 | 85.3 | 72.4# | exhaustiveness=64 |
| DiffDock | 63.0 | 84.8 | 55.9# | - |
| DSDP (shape) | 57.2 ± 1.5 | 72.6 ± 1.7 | 0.82# | 384 × 40 |